\documentclass[tradiabstract]{aa} 
\usepackage{graphicx}
\usepackage{longtable}
\usepackage{txfonts}
\begin{document}
\title{The Arches cluster revisited. III. An addendum to the stellar census
\thanks{Based on observations made at the European Southern Observatory, 
Paranal, Chile under programmes ESO 087.D-0317, 091.D-0187, 093.D-0306, 099.D-0345 and
0101.D-0141 } 
}
\author{J.~S.~Clark\inst{1}
\and M.~E.~Lohr\inst{1}
\and L.~R.~Patrick\inst{2,3}
\and F.~Najarro\inst{4}}
\institute{
$^1$School of physical sciences, The Open 
University, Walton Hall, Milton Keynes, MK7 6AA, United Kingdom\\
$^2$Instituto de Astrof\'{\i}sica de Canarias, E-38205 La Laguna, Tenerife, Spain\\ 
$^3$Universidad de La Laguna, Dpto Astrof\'{i}sica, E-38206 La Laguna, Tenerife,
Spain\\
$^4$Departamento de Astrof\'{\i}sica, Centro de Astrobiolog\'{\i}a,
(CSIC-INTA), Ctra. Torrej\'on a Ajalvir, km 4,  28850 Torrej\'on de Ardoz,
Madrid, Spain}

   \abstract{The Arches is one of the youngest, densest and most massive clusters in the Galaxy. 
As such it provides a unique insight into the lifecycle of the most massive stars known and the formation and  survival of such stellar aggregates in the extreme conditions of the Galactic Centre. In a previous study we presented an initial stellar census for the Arches and in this work we expand upon this, providing new and revised classifications for $\sim30$\% of the 
105 spectroscopically identified cluster members as well as distinguishing  potential massive runaways. The results of this survey emphasise the homogeneity and  co-evality of the Arches and confirm the absence of H-free Wolf-Rayets of WC sub-type and predicted luminosities.
The increased depth of our complete dataset also provides significantly better constraints on the main sequence population; with the identification of 
O9.5 V stars for the first time we now spectroscopically sample stars with initial masses 
ranging from $\sim16M_{\odot}$ to $\geq120M_{\odot}$.  Indeed, following from our expanded stellar census  we might expect 
$\gtrsim50$ stars within the Arches to have been born with masses $\gtrsim60M_{\odot}$, while all 105 spectroscopically confirmed 
cluster members are massive enough to leave relativistic remnants upon their demise.
 Moreover the well defined observational properties of the main sequence cohort will be 
critical to the construction of an extinction law appropriate for the Galactic Centre and consequently the quantitative analysis of the Arches population and subsequent determination of the cluster initial mass function.    }

\maketitle

\section{Introduction}

Located within the central molecular zone (CMZ) of the Galaxy at a projected distance of $\sim30$pc from SgrA$^*$,
the Arches is one of the most extreme young massive clusters in the Milky Way in terms of youth, integrated mass and stellar density. Independently discovered by Nagata et al. (\cite{nagata}) and Cotera et al. (\cite{cotera96}), it has since become 
a lodestone for understanding the formation of very massive stars and starburst clusters in extreme conditions; the survival, 
structure and evolution of such aggregates in the gravitational potential of the Galactic centre; the form of the initial mass function and in particular the presence or otherwise of a high-mass cut-off; and  the evolution and final end-points of the most massive stars nature permits to form in the local Universe. 

 Given this potential, much effort has been expended on constraining the form of the (initial) mass function via deep, high 
 angular resolution photometric observations  (e.g. Figer et al. \cite{figer99}, Stolte et al. \cite{stolte02}, Kim et al. 
\cite{kim}, Espinoza et al. \cite{espinoza}, Clarkson et al. \cite{clarkson}, Habibi et al. \cite{habibi}
Shin \& Kim \cite{shin} and Hosek et al. \cite{hosek}). However such efforts  are compromised by significant uncertainties in the calibration of the mass/near-IR luminosity function, due to uncertainties over the form of the interstellar extinction law to apply  which, in turn, yield errors of up to $\sim0.6$dex in the bolometric luminosity of cluster members (Clark et al. \cite{clark18a}; henceforth Paper I). 

In parallel, multiple near-IR  spectroscopic observations have been undertaken in order to better understand the constituent stars, construct a cluster   HR diagram and hence infer fundamental properties such as cluster age and the masses of  individual members (Figer et al. \cite{figer02}, Najarro et al. \cite{najarro}, Martins et al. \cite{martins08}). Again prone to uncertainties regarding interstellar extinction, these studies have hinted at an extended formation history for cluster members, although Schneider et al. 
(\cite{schneider}) suggest this could instead result from binary interaction and rejuvenation. 

Motivated by these efforts we undertook a multi-epoch spectroscopic study of the Arches in order to determine the properties of the binary population and, via the  summation of multiple individual   observations, obtain high S/N spectra for the production of a stellar census, derivation of an extinction law and the  quantitative determination of  stellar and cluster parameters. Our initial census (Paper I)
was based on three VLT/SINFONI observing runs in 2011, 2013 and 2017 
that each remained  substantially incomplete - a total of 16.4hr of observations were formally 
completed, with a  further 10.1hr observations undertaken in non-optimal conditions. 
After acceptance of Paper I, a  further 12.4hr and 1.8hr of observations were executed in optimal and non-optimal conditions between 
2018 April-August. Additional multi-epoch VLT/KMOS observations were also made of a 
handful of brighter cluster members\footnote{Stars B1, F1, F6, 
F8, F9 and F16.} as part of  a wider survey of  massive stars distributed thoughout the 
CMZ (Clark et al. \cite{clark18c}); a number of stars in the 
vicinity of the Arches that had been flagged as potential runaways by Mauerhan et al. (\cite{mauerhan}) 
were also observed as part of this effort. 

In this short  supplemental paper we present the results from both  observational programmes which, when 
combined with extant data, substantially improves on cluster completeness and yields much improved S/N 
spectra for  a large number of cluster members, enabling more robust spectral classifications and quantitative
analysis. The results of our radial velocity survey will be presented in a companion
  paper (Lohr et al. \cite{lohr19}). 

\begin{figure*}
\includegraphics[width=12cm,angle=0]{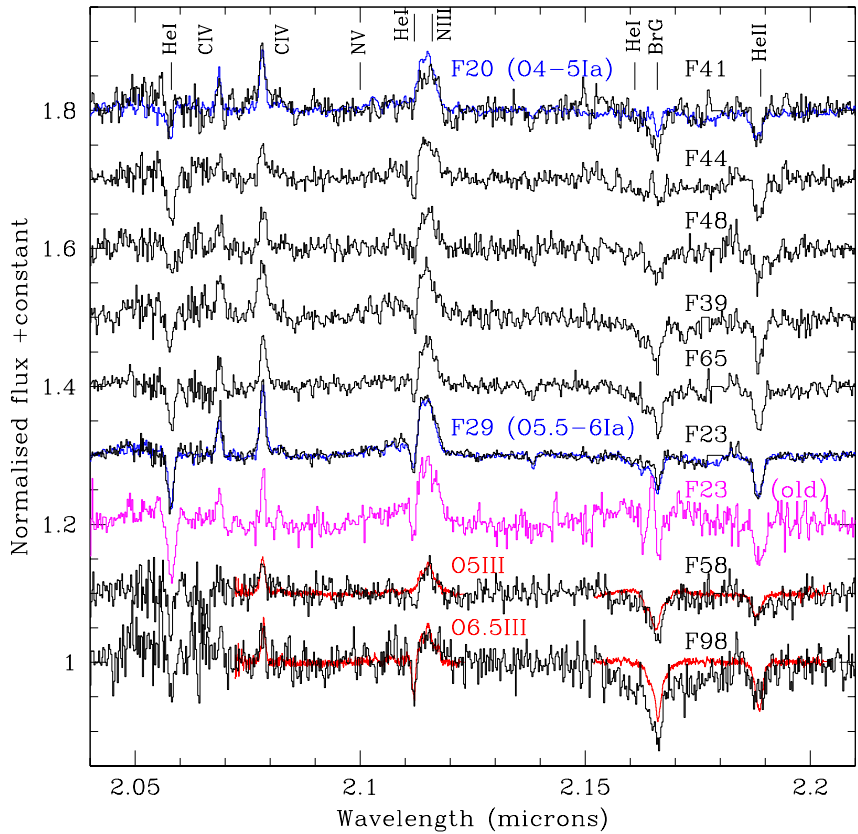}
\caption{K-band spectra  of new or reclassified O4-6 giant and supergiant stars within the Arches. Spectra of
supergiant cluster members  F20 and F29 are shown for comparison (blue) as are the template spectra for O5 
III (HD 15558) and O6.5 III  (HD 190864) stars (red; Hanson et al. \cite{hanson05}). Finally the spectrum of
F23 from Paper I is reproduced  (magenta) in order to illustrate the increased  S/N 
in  the new data and also the removal of spurious emission in the Br$\gamma$ profile of this particular star.}
\end{figure*}

\begin{figure*}
\includegraphics[width=12cm,angle=0]{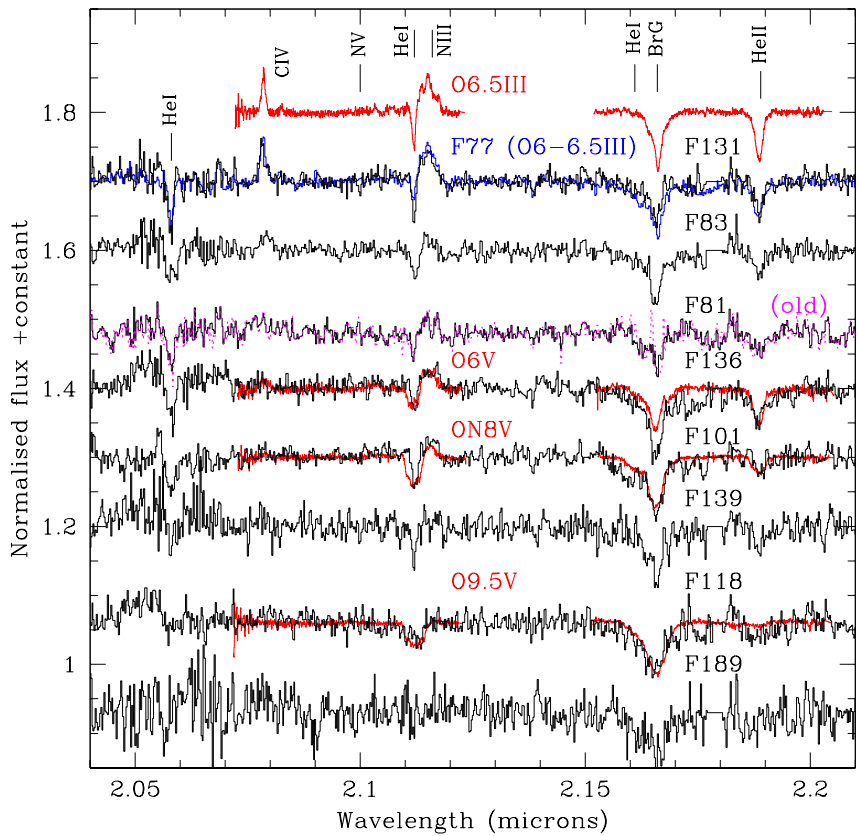}
\caption{K-band spectra of new  or reclassified O6-O9.5 main sequence and giant stars within the Arches. A
spectrum  of the  O6-6.5 III cluster  member   F77 is shown for comparison (blue) as are the template spectra
for O6.5 III (HD 190864), O6 V (HD 5689), ON8 V (HD 13268)  and O9.5 V (HD 149757)  stars (red; Hanson et al.
\cite{hanson05}). Finally the spectrum of F81  from Paper I is reproduced (magenta) in order to illustrate
the removal of spurious emission in the Br$\gamma$ profile.}
\end{figure*}

\begin{figure*}
\includegraphics[width=13.2cm,angle=0]{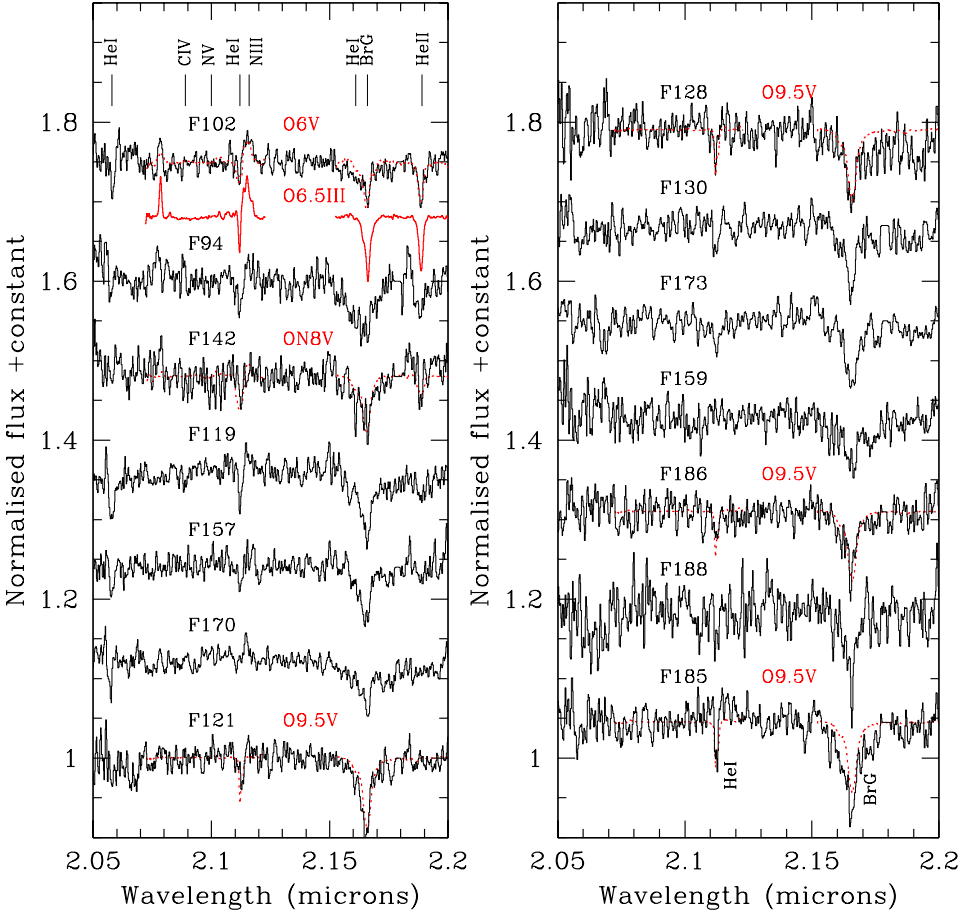}
\caption{K-band spectra of additional giant and  main sequence cluster members; these have been rebinned to
improve the  S/N ratio. Comparison spectra for O6 V (HD 5689), O6.5 III (HD 190864), ON8 V (HD 13268) and
O9.5 V (HD 37468) are overplotted in red and from Hanson et al. (\cite{hanson05}).}
\end{figure*}

\begin{figure*}
\includegraphics[width=9.1cm,angle=-90]{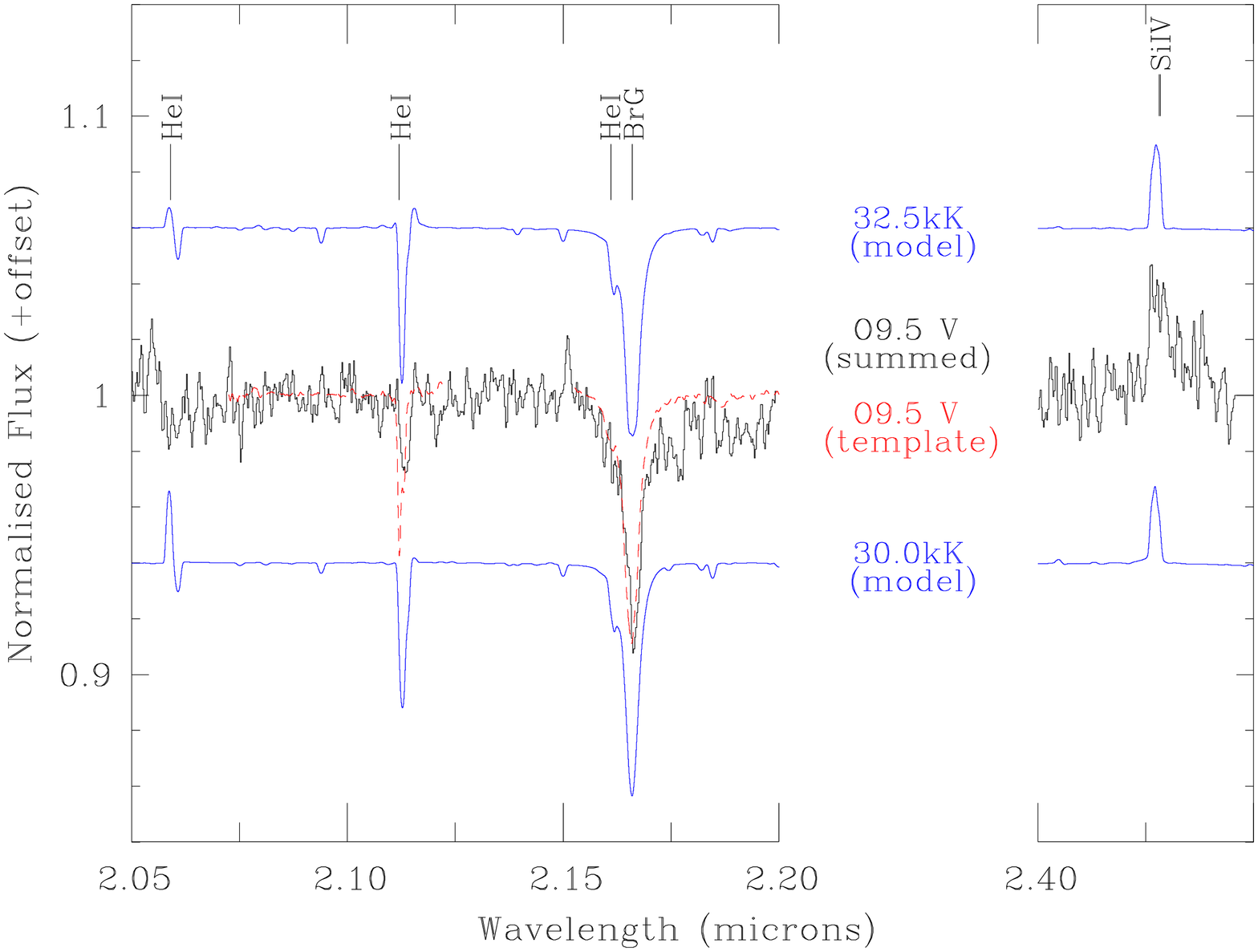}
\caption{K-band spectrum derived from the combination of individual spectra of candidate O9.5 V stars (black) 
plotted against template O9.5 V spectrum (HD 37468) and synthetic spectra for 32.5kK and 30.0kK O dwarfs 
(blue; Sect. 3.1). The region of the spectrum between 2.2$\mu$m and 2.4$\mu$m is both predicted and found to be featureless.} 
\end{figure*}

\section{Data acquisition, reduction and analysis}

Data acquisition and reduction for the VLT/SINFONI and VLT/KMOS observations closely followed the 
methodologies described in Clark et al. (\cite{clark18a}, \cite{clark18b}). The sole revision 
was in field placement for the SINFONI observations, which sampled a number of previously 
unobserved fields in the periphery of the cluster. These observations were undertaken between 
the nights of 2018 May 6 and August 23.

A total of 78 spectra were extracted from the VLT/SINFONI observations undertaken in 2018. Of these one is a previously unobserved cool star interloper (F36), which is not discussed further\footnote{For completeness F11, F36, F46, F51 and F99 are all interlopers.} and two  are of  spatially unresolvable blends of two cluster members, leading to spectra of 75 unique objects. Ten of these are the first observations of individual stars. After combination with extant data and the rebinning of 15 spectra to improve their  S/N
(at a cost of spectral resolution) the resultant dataset permitted analysis of a further  eight previously unclassifiable objects 
(due to  prior low S/N  (3 stars) or potential blending (5 stars, see below)), with a revision  of existing spectral type or luminosity class estimates possible for an additional 14 stars (Table 1). 
The remaining  44 spectra are  consistent with current classifications (Paper I) but are of improved S/N. These data are still 
valuable since the greater S/N  better constrains the shape of line profiles that are utilised in quantitative model-atmosphere 
analysis - an example being Br$\gamma$ which is sensitive to wind properties. 

 We also took the opportunity to revisit objects that had previously been considered blends (Paper I).  A number of stars are 
 effectively merged with much brighter companions and hence are inaccessible\footnote{F31, F37, F66, F70, F76, F109, F122, F123, 
 F140 and F167 are  merged with F20, F7, F7, F1, F12, F7, F12, F28, F18,  and F35, respectively.}. F56+F95 and F57+72 are pairs of 
 stars of comparble brightness that are fully blended and so we are only able to present composite spectra in order to demonstrate 
 that the stars they derive from are not atypical when compared to the wider population. The blended F124 and F143 are likewise of 
 comparable luminosity but are sufficiently faint to not justify extraction; as is F169, which is merged with a number of anonymous  faint stars.

 Inspection of the individual spectra of the previously unclassified stars F44 (close to F73), F73, F80 (close to F7), F142 (close 
 to F53) and F170 (close to F32, F33 and F38) suggest that they are unlikely to be contaminated since the stellar features in their spectra differ in presence, strength and profile from their near neighbours. As a consequence we provide the 
 first classifications  for these objects (Sect. 3.1) but highlight their proximity to brighter neighbours in Table 1.   No new 
 observations were made of F166 (paper I); however its mid-O III-V star appearance is unexpected given its comparative faintness; 
 hence we consider it likely that it is likely to be contaminated  by its bright O hypergiant neighbour F27 and do not consider it 
 further.

\section{Data presentation, spectral classification and interpretation}

As with data extraction and reduction, spectral classification followed the methodology outlined in Paper I. Spectra of the subset of cluster members for which categorisation - or revision of existing classifications - is now possible are presented in Figs. 1-3. Fig. 4 presents a composite spectrum derived from multiple low S/N spectra of late-O main sequence candidates in comparison to template and synthetic classification spectra (Sect. 3.1), while in Fig. 5 we plot spectra of two unresolvable blends and a further four anomalous spectra. Finally, the spectra  of  potential runaways are shown in  Fig. 6. 

\subsection{Cluster members}

We present the first observations of F39, 41  and F48 in Fig. 1, finding them all to be new mid O supergiants (Table 1), as is 
the previously unclassified F44. F65 is likewise  revised from luminosity class I-III to Ia. The spectral type of F23 is 
marginally earlier than prevously reported but  importantly the strong narrow emission peak superimposed on the  Br$\gamma$ 
photospheric profile is found to be spurious  (Fig. 1 and Table 1), as are similar features  in F81 and F139 (Fig. 2 and Paper I). This is 
important as the shape and strength of this  feature is an important diagnostic of stellar luminosity and wind properties; hence 
while the improved S/N of the new spectra of  other cluster members may not affect their classification they will aid 
quantitative analysis. 

Comparison of the previously unobserved stars F58 and F98 to classification templates reveals both to be mid-O giants (Fig. 1 and Table 1, although note the anomalously strong Br$\gamma$ profile of F98) as are the previously unclassified F73 and F80 (Paper I). The much improved S/N of F131 allows a significant reclassification from $>$O8 V (Paper I) to O6.5 III (Fig. 2); surprising given its relative faintness in comparison to other giants. Despite  proximity to both F10 and F17, inspection of their spectra suggests that they are unlikely to contribute to that of F131. Instead consideration of the IR colours of F131 and its spectroscopic twin F77 (Fig. 2) reveals the former to be redder, suggesting that it suffers  anomalously high interstellar extinction\footnote{We highlight that F131 is located in the SW of the cluster; a region apparently subject to 
unexpectedly high reddening (cf. results for F2; Lohr et al. \cite{lohr18a}).}.
Weaker emission in C\,{\sc iv}  2.079$\mu$m and the 2.11$\mu$m blend implies a slightly later luminosity type for the newly 
observed F83 (O6-7 III-V; Fig. 2), bolstered by its similarity to F81. Finally, despite a low S/N which necessitated rebinning, 
the new spectrum of F94 is also consistent  with such an identification (Fig. 3; noting the anomalously broad and strong 
Br$\gamma$ profile).

 Next we turn to the main sequence (MS) cohort (Figs. 2 and 3) where we provide the first classifications for  eight stars and 
reclassifications for a further 11.  Despite its comparatively low S/N even after rebinning, the presence of prominent C\,{\sc iv} 
2.089$\mu$m emission and He\,{\sc ii}  2.189$\mu$m absorption in the spectrum of F102 (Fig. 3) allows a refined classification of 
O5-6 V; the earliest spectral type sampled here. The strength of the He\,{\sc i} 2.112$\mu$m and He\,{\sc ii} 2.189$\mu$m 
temperature diagnostics  suggest an O6 V classification for F136 although Br$\gamma$ is anomalously strong and we see no evidence 
of (weak) C\,{\sc iv}  2.089$\mu$m emission; we adopt an initial  (conservative) classification of O6-7 V. Of the remaining stars 
from Fig. 2, the presence of weak He\,{\sc ii} 2.189$\mu$m absorption in F101 and F139 indicates that both are O8 V stars. 
The absence of this line  and  the strength of  Br$\gamma$ is consistent with  an O9.5 V classification for F189, although  the lack of  O8.5 V and O9 V template spectra for comparison leaves open the possibility of a slightly earlier spectral type for this and other similarly classified stars. Intriguingly, 
the S/N of the F118 spectrum is sufficiently high to reveal an excellent coincidence between its He\,{\sc i} 2.112$\mu$m profile 
and that of the  high {\em v}sin{\em i}  O9.5 V template star HD 149757 (Hanson et al. \cite{hanson05}).  One would  also 
expect  broadening in the  He\,{\sc ii} 2.189$\mu$m line in rapidly rotating stars and, given its comparative  weakness in 
mid-late O stars, one might  anticipate it being  undetectable in relatively low S/N spectra, erroneously leading to the assignment of slightly later spectral types.  Indeed rebinning the spectrum of F118 
to  improve the S/N suggests a possible detection of this feature; hence we assign an O8-9.5 V classification to this star. 

As with F101 and F139 (Fig. 2), the strength of He\,{\sc ii} 2.189$\mu$m and Br$\gamma$ absorption, when combined with a lack  of 
C\,{\sc iv} 2.089$\mu$m emission, suggests that F119 and F142 (Fig. 3) are both O8 V stars. A marginal detection 
of He\,{\sc ii} 2.189$\mu$m in F157 results in an O8-O9.5 V classification. While this line appears absent in F170, the Br$\gamma$ 
line appears unexpectedly weak for an O9.5 V star (Fig. 3); hence we  likewise adopt O8-O9.5 V for it. All the remaining stars in 
Fig. 3 lack  He\,{\sc ii} 2.189$\mu$m absorption and appear broadly consistent with O9.5 V classifications (subject to the {\em caveats} above), with the greater depth of the 
Br$\gamma$  profile in F185 hinting at a still later spectral type. He\,{\sc i} 2.112$\mu$m absorption - present until spectral 
type B2-3 (e.g. Clark \& Steele \cite{clark00}) - appears absent in some sources (e.g. F159); we suspect this may be due to a 
combination of low S/N and an intrinsically  broad, shallow line profile due to rapid rotation (c.f. F118).

In order to verify these classifications we summed the spectra of our low-S/N O9.5 V candidates\footnote{F121, F128, F130, F159, F173, F186 and F188.} to produce a single composite spectrum of higher S/N\footnote{
Excluding the significant overheads associated with telescope preset, instrumental set-up, target acquisition and telluric 
observations for the multiple observations used to construct the composite spectrum, but including sky observations and overheads associated with read-out, this amounts to a total of $\sim4.6$hr 
of observations, of which time-on-target is 60\% of this total.}.
This is reproduced in Fig. 4 where an excellent 
correspondence is  found with the O9.5 V template spectrum (Hanson et al. \cite{hanson05}).  Foreshadowing future quantitative 
analysis and to  further test our conclusions we also present {\em illustrative} synthetic spectra of O dwarfs generated  with 
the non-LTE model-atmosphere code  CMFGEN (Hillier \& Miller \cite{hm1}, \cite{hm2}). We caution that  we have not 
attempted to  specifically fit the composite spectrum, adopting canonical  effective temperatures for O8.5 V  ($T_{\rm 
eff}\sim32.5$kK) and  O9.5 V stars ($T_{\rm eff}\sim30.0$kK;  following the calibration of Martins et al. \cite{martins05}). A 
surface gravity of log$g\sim4.0$ was employed and solar metalicity was assumed for the iron group elements,
while we used a factor of two enhancement for  $\alpha$-elements  with respect to iron (Najarro et al. \cite{paco09}).
Finally the resultant spectra were convolved with an assumed {\em v}sin{\em i}$\sim175$kms$^{-1}$ and degraded to the resolution 
of our observations. 

The correspondence of both synthetic spectra with our composite spectrum is encouraging (Fig. 4). The He\,{\sc i} 2.059$\mu$m line 
is highly sensitive to the UV radiation field and while the He\,{\sc i} 2.112$\mu$m photospheric line appears a little strong, it is 
sensitive to both the  turbulent velocity ($v_{\rm turb}$) and rotational broadening. However we find an excellent `fit' to the 
He\,{\sc i}+Br$\gamma$ photospheric blend in 
terms of both strength and  line profile. Moreover our models predict emission in the Si\,{\sc iv} $\sim2.427\mu$m blend, which is 
an excellent hot star  diagnostic (cf. Clark et al. \cite{clark18b}). This line is clearly in emission in our composite spectrum,  with the unexpected breadth of the line likely  due to incomplete correction of the significant telluric contamination that plagues this wavelength region.  Given the consonance with both template 
and synthetic spectra we conclude that the  individual component stars are indeed late-O dwarfs, with temperatures between 
30-32kK (footnote  5 and Table 1); assigning primacy to the template rather than synthetic spectra we classify these stars as O9.5 V.

 Finally we turn to the anomalous spectra presented in Fig. 5. The blended spectra of F56+95 and F57+72 appear to resemble 
 mid-O giants; as a consequence it appears unlikely that any of the constituent stars differ greatly from the wider 
 population. As to F92 and F93, the presence of  emission in the He\,{\sc i}/N\,{\sc iii}/O\,{\sc iii}/C\,{\sc iii} 2.11$\mu$m 
 blend and  C\,{\sc iv} 2.079$\mu$m line, when combined with  He\,{\sc i} 2.059$\mu$m absorption (and for F92 also  He\,{\sc i} 
 2.112$\mu$m) is likewise consistent with a mid-O classification (Table 1). However the absence of pronounced 
  Br$\gamma$ and He\,{\sc  ii} 2.189$\mu$m photospheric lines is unexpected. Some trace of the latter may be evident in both 
 spectra; the same may be true for Br$\gamma$ in F93, while weak, broad emission may be present in F92.

Motivated by these 
findings we examined the longer wavelength portion of both spectra. CO bandhead emission is absent for both sources but  Si\,{\sc 
iv} $\sim2.427\mu$m is present, confirming both are early-type stars. 
Additionally weak  N\,{\sc iii} 2.247$\mu$m and 2.251$\mu$m emission is present in  F92; unfortunately the S/N of F93 is too poor 
for conclusions to be drawn. The combination of Si\,{\sc iv} and N\,{\sc iii} emission is seen in mid O giants such as 
F77, corroborating the classifications derived from shorter wavelength diagnostics. We speculate  that excess 
 emission from unexpectedly strong stellar winds may result in the infilling of the  Br$\gamma$ and He\,{\sc  ii} 2.189$\mu$m 
photospheric lines for both stars; the lack of CO bandhead emission arguing against a contribution from a cool circumstellar disc 
(such as found for a subset of  low luminosity cluster members; Stolte et al. \cite{stolte10}, \cite{stolte15}).

The spectra of F150 and F151 appear to represent even more extreme examples of this phenomena 
- emission in Br$\gamma$ and the 2.11$\mu$m blend (with He\,{\sc i} 2.112$\mu$m absorption)
is present in the former while the latter is entirely featureless. Again CO bandhead emission is absent 
from these stars while the low S/N of the spectra in this region preclude conclusive 
identification of Si\,{\sc iv} emission.   Noting that both stars are magnitudes fainter than the 
dust-enshrouded WCL stars found  within the Quintuplet (Paper I, Clark et al. \cite{clark18b}) we 
are unable to offer an  evolutionary  classification for these objects, but simply assume that 
the spectra are dominated by emission from circumstellar material.

\begin{figure*}
\includegraphics[width=9.1cm,angle=-90]{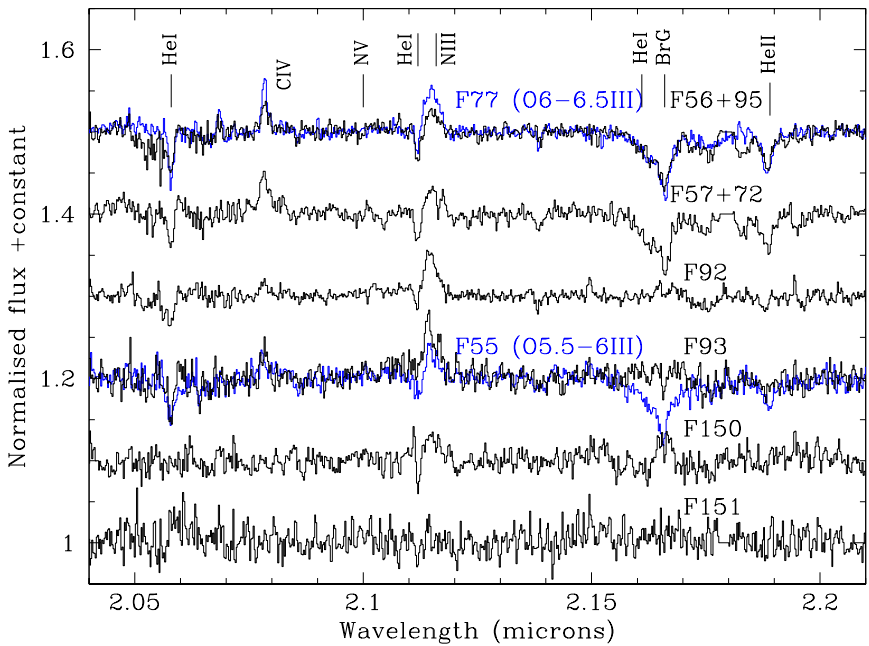}
\caption{K-band  spectra of the blends of F56+F95 and F57+72 along with the anomalous souces F92, F93, F150 
and F151 (see
Sect. 3.1). Spectra of the cluster members F55 and F77 are overplotted in blue for comparison.}
\end{figure*}

\begin{figure*}
\includegraphics[width=9.1cm,angle=-90]{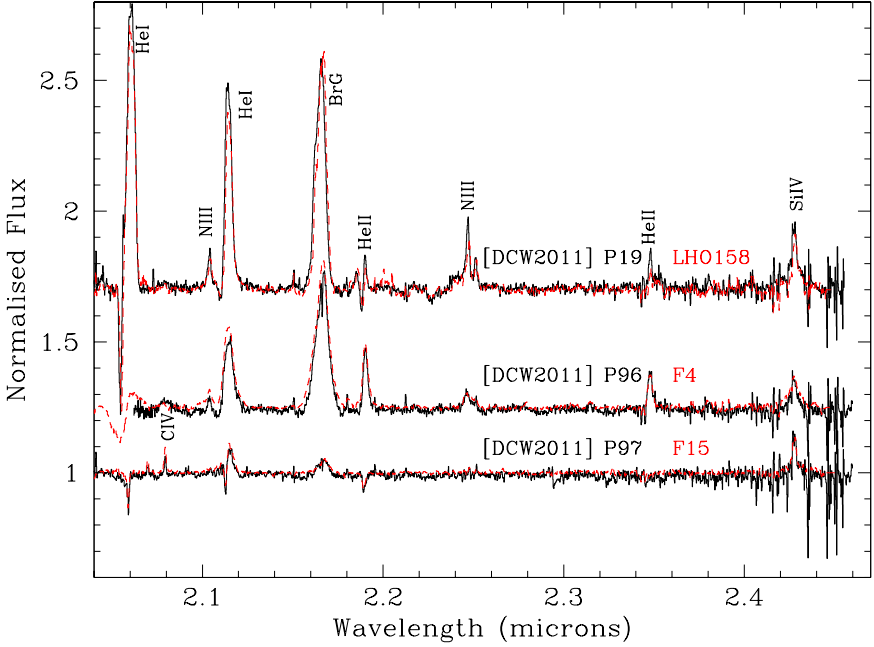}
\caption{K-band spectra  of candidate runaways (black) compared to members of the Arches and Quintuplet (red; 
Clark et al.
\cite{clark18a},\cite{clark18b}).  [DCW2011] P96 and P97 show a close resemblance to WN7-8h and O6-7 Ia$^+$ 
stars within the
Arches, while [DCW2011] P19 corresponds to the sole broad-lined WN9h star within the older Quintuplet.  }
\end{figure*}

\begin{table*}
\caption{The expanded stellar population of the Arches cluster}
\begin{center}
\begin{tabular}{lccccccccl}
\hline
\hline
ID &        RA     &       Dec     &  $m_{\rm F127M}$   & $m_{\rm F139M}$   &   $m_{\rm F153M}$  & $m_{\rm F205W}$ & \#Observations & 
Spectral & Notes \\
   &    (h m s)    &     (d m s)   &     (mag)  & (mag)     &   (mag)       &  (mag)     & (\#Epochs)   & Classification &       \\
\hline
F23&  17 45 51.211 &  -28 49 23.84 &  16.38     &  15.27     &   14.21    & 12.19 & 6(6) & O5.5-6 Ia &  \\
F39&  17 45 51.166 &  -28 49 36.70 &  16.98     &  15.84     &   14.76    & 12.65 & 1(1) & O5.5-6 Ia &  \\
F41&  17 45 50.429 &  -28 49 27.42 &  18.46     &  17.15     &   15.89    & 13.53 & 4(4) & O5 Ia     &  \\
F44&  17 45 50.701 &  -28 49 25.59 &  17.19     &  16.05     &   14.93    & 12.88 & 7(5) & O5.5-6 Ia & Close to 73 \\
F48&  17 45 50.623 &  -28 49 26.96 &  17.69     &  16.53     &   15.42    & 13.28 & 6(6) & O5.5-6 Ia &  \\
F56&  17 45 50.591 &  -28 49 22.19 &  17.00     &  15.92     &   14.91    & 13.03 & 2(2) & {\em O5.5-6 III}&  Merged with F95  \\
F57&  17 45 50.768 &  -28 49 20.31 &  16.91     &   -        &     -      & 13.04 & 2(2) & {\em O5.5-6 III}&  Merged with F72 \\
F58&  17 45 49.968 &  -28 49 19.60 &  17.83     &  16.57     &   15.37    & 13.05 & 3(3) & O5.5-6 III&  \\
F65&  17 45 50.082 &  -28 49 21.53 &  17.58     &  16.42     &   15.32    & 13.16 & 5(5) & O5.5-6 Ia &  \\
F72&  17 45 50.784 &  -28 49 20.27 &    -       &  16.08     &   15.05    & 13.13 & 2(2) & {\em O5.5-6 III}&  Merged with F57 \\
F73&  17 45 50.679 &  -28 49 25.58 &  17.78     &  16.59     &   15.48    & 13.35 & 11(9) & O6 III    & Close to F44 \\
F80&  17 45 50.589 &  -28 49 19.46 &  17.40     &  16.29     &   15.32    & 13.47 & 28(14) & O6-6.5 III  & Close to F7  \\
F83&  17 45 50.061 &  -28 49 20.67 &  18.00     &  16.81     &   15.69    & 13.53 & 4(4) & O6-7 III-V &  \\
F89&  17 45 50.955 &  -28 49 17.15 &  17.33     &  16.33     &   15.39    & 13.65 & 16(15) & O7-8 V    &  \\
F94&  17 45 50.657 &  -28 49 28.03 &  18.74     &  17.50     &   16.33    & 14.14 & 4(4) & O6-8 III-V  &  \\
F95&  17 45 50.578 &  -28 49 22.00 &  17.80     &  16.71     &   15.69    & 13.71 & 2(2) & {\em O5.5-6 III}&  Merged with F56 \\
F98&  17 45 51.153 &  -28 49 37.06 &  18.08     &  16.93     &   15.87    & 13.75 & 1(1) & O6-6.5 III    &  \\
F101& 17 45 50.914 &  -28 49 18.37 &  17.55     &  16.53     &   15.58    & 13.78 & 19(16) & O8 V      &   \\
F102& 17 45 50.876 &  -28 49 28.80 &  18.24     &  17.07     &   15.93    & 13.78 & 6(6) & O5-6 V      &  \\
F118& 17 45 50.453 &  -28 49 25.97 &  18.66     &  17.44     &   16.28    & 14.08 & 8(6) & O8-9.5 V    &  \\
F119& 17 45 50.842 &  -28 49 22.75 &  18.18     &  17.09     &   16.06    & 14.06 & 11(10) & O8 V      & \\ 
F121& 17 45 50.204 &  -28 49 19.64 &  18.53     &  17.35     &   16.24    & 14.09 & 9(9) & O9.5 V &  \\
F128& 17 45 51.216 &  -28 49 36.53 &  18.51     &  17.37     &   16.28    & 14.18 & 1(1) & O9.5 V &   \\
F130& 17 45 50.605 &  -28 49 27.50 &  18.80     &  17.58     &   16.40    & 14.21 & 4(4) & O9.5 V    &  \\
F131& 17 45 50.163 &  -28 49 27.46 &  19.11     &  17.80     &   16.55    & 14.23 & 9(6) & O6.5 III  & Close to F10 \& 17 \\
F136& 17 45 50.077 &  -28 49 27.84 &  19.59     &  18.18     &   16.83    & 14.30 & 13(10) & O6-7 V      &  \\
F139& 17 45 51.207 &  -28 49 23.41 &  18.59     &  17.48     &   16.33    & 14.36 & 6(6) & O8 V      &  \\
F142& 17 45 50.689 &  -28 49 24.93 &  18.63     &  17.50     &   16.43    & 14.41 & 2(2) & O8 V      & Close to F53 \\
F157& 17 45 50.220 &  -28 49 24.18 &  19.11     &  17.89     &   16.71    & 14.55 & 14(14) & O8-9.5 V &  \\
F159& 17 45 50.882 &  -28 49 15.69 &  19.63     &  18.49     &   17.32    & 14.57 & 6(6) & O9.5 V    & \\
F170& 17 45 50.704 &  -28 49 20.96 &  18.74     &  17.67     &   16.63    & 14.74 & 19(10) & O8-9.5  V & Close to F32, 33 \& 38 \\
F173& 17 45 50.204 &  -28 49 20.24 &  19.10     &  17.93     &   16.83    & 14.80 & 5(5) & O9.5 V    &  \\
F185& 17 45 50.282 &  -28 49 27.56 &  19.65     &  18.40     &   17.18    & 14.95 & 10(7) & O9.5 V    &  \\
F186& 17 45 50.001 &  -28 49 28.24 &  20.43     &  19.01     &   17.61    & 14.95 & 13(10) & O9.5 V    &  \\
F188& 17 45 50.936 &  -28 49 21.80 &  19.02     &  17.93     &   16.91    & 14.97 & 3(3) & O9.5 V    &  \\
F189& 17 45 50.595 &  -28 49 23.52 &  18.91     &  17.79     &   16.69    & 14.77 & 5(5) & O9.5 V    &  \\
\hline
\end{tabular}
\end{center}
{We summarise the properties of  stars with new or revised spectral classifications. Column  1 indicates the nomenclature for cluster members adopted  by Figer et al. (\cite{figer02}), columns 2 and 3 the J2000 co-ordinates, columns 4-6 the HST WFC3 photometry from Paper I and column 7  F205W filter photometry from Figer et al. (\cite{figer02}). Column 8 presents the  total number of VLT/SINFONI data-cubes available 
for individual objects, with the number in parentheses being the number of epochs on which these data were obtained. Column 9 provides a spectral classification while the final column provides additional notes on possible blending or contamination; individual cases are discussed in Sect. 2. F36 is found to be a cool interloper; as with F11, F46, F51 and F99 it is therefore excluded from the table. Finally the nominal classification of the unresolvable blends F56+F95 and F57+72 are given in italics.}
\end{table*}

\subsection{Runaways}

Mauerhan et al. (\cite{mauerhan}) discuss three stars identified by the Pa$\alpha$ survey of Dong et al (\cite{dong11}) 
as potential runaways -  [DCW2011] P19, P22  and P96 (=\#11 and \#12 and G0.10+0.02, respectively, in the former work) - 
which are all located  at a projected distance of 1-2pc from the Arches. To these we may add a fourth, 
[DCW2011] P97, at a slightly larger distance of 
4-5pc (Dong et al. \cite{dong15}). 

 Comparison of our spectrum of [DCW2011] P19 to that of the Quintuplet cluster member LHO158 (Fig. 6) shows a striking 
resemblance, slightly refining the spectral classification to that of an anomalously 
 broad-lined WN9h star. Such stars likely have lower temperatures and mass-loss rates than the least extreme  
WN7-9ha stars within the Arches (which at 2-3Myr is younger than the 3-3.6Myr old Quintuplet; Paper I,
Clark et al.  \cite{clark18b}); hence we consider its physical association with the cluster  unproven at this time.  

In contrast Mauerhan et al. (\cite{mauerhan}) classify [DCW2011] P22 as WN8-9h; comparison of their spectrum to the WNLh  cohort 
in the 
Arches shows a close correspondence to the least extreme example, F16. The only previous spectra of [DCW2011] P96 were of low 
 resolution and S/N (Cotera et al. \cite{cotera99}); our new observation shows it to be a doppleganger of the WN7-8h star F4 
 (Fig. 6) and we adopt this as a classification. Finally Dong et al. (\cite{dong15}) suggest O4-6 Ia$^+$ for [DCW2011] P97; on 
the basis of its marked similarity to F15 we slightly revise this to O6-7 Ia$^+$.

Considering photometric observations and  the $H-$ and $K-$band magnitudes of [DCW2011] P22 and P97 deviate from their Arches comparators by only 0.1mag in the 
$K-$band  (Dong et al. \cite{dong11}).  [DCW2011] P96 is significantly fainter than F4 but is also redder, suggesting differential 
extinction  may cause this discrepancy; in any event it lies within the envelope of magnitudes exhibited by the population of WNLh 
stars within  the Arches. 

Assuming a runaway velocity of $\sim10$km$^{-1}$ it would take only $\sim98,000$yr ($\sim244,000$yr) to 
travel a projected  distance of 2pc (5pc) from the Arches, significantly less than the age of the cluster ($\sim2-3$Myr; Paper 
I). Given this and  the close equivalence of both spectroscopic  and photometric properties of [DCW2011] P22, P67 and P97
to Arches  members, we consider it likely that they are indeed runaways from the cluster; either as a result of dynamical 
interactions  (e.g. Poveda et al. \cite{poveda}) or tidal stripping/disruption (e.g. Habibi et al. \cite{habibi14}, Park et al. 
\cite{park}).  Moreover it is unlikely that these stars comprise the complete runaway population; other stars of comparable 
spectral type  are located at larger distances throughout the CMZ, while the O supergiants (and less evolved stars) found within 
the Arches  support  weaker winds and hence will not have been detected by the Pa$\alpha$ survey of Dong et al. (\cite{dong11}).

\section{Discussion and concluding remarks}

\subsection{Observational completeness}

At this juncture it is worth considering the completeness of the current sample. We are
now able to provide spectral classifications for a total of  105 cluster members.  This includes the 
first classifications for 18 stars and the reclassification of a further 14 objects, but  excludes two 
unclassifiable stars (F150 and F151), the unresolvable blends F56+F95 and F57+72 and 
the three putative runaways (Sect. 3.2). Stars F1-F51, B1 and B4 have all been observed multiple times, and have 
either been identified as cluster members (47 objects; Paper I and Table 1), rejected as non-members (F11, 36, 46 and 51) or were 
too close to other, brighter stars to extract a useful spectrum (F31 and 37). With the  exception of F41 and F50, these correspond to stars with $m_{\rm F205W}\sim10.4-12.9$ (Figer et al. \cite{figer02}). Excluding blends and interlopers (seven objects; Sect. 2), in the range F52-F100 ($m_{\rm F205W}\sim12.9-13.8$) 28 stars have spectra of sufficient quality to allow classification as {\em bona fide} cluster members. Thus, of the  102 brightest stars potentially associated with the Arches we identify 75 as constituents, with only 13 objects remaining to be observed\footnote{F52, F59, F61, F67, F71, F75, F78, F79, F86, F88, F91, F97 and  F100.}.

 Of the remaining 96 stars identified by Figer et al. (\cite{figer02}; F101-196 ($m_{\rm F205W}\sim13.8-15.0$)) observations of sufficient S/N for classification have been made for 29 objects (Paper I and Table 1). Spectra of five stars were of insufficient S/N to permit this\footnote{F166, F168, F174, F176 and F184.}, while  a further four objects were judged too faint to make extraction worthwhile\footnote{F158, F163, F182 and  187.}. Of the remainder, seven stars are inaccessible due to blending (Sect. 2) leaving 48 that are still to be observed\footnote{F103-109, F111, F113, F116, F120, F125-127, F129, F132-134, F137, F138, F141, F144-149, F152, F154, F156, F160-162, F164, F165, F171, F175, F178-181, F183, and F190-196.}; however it appears likely that many of these will likewise prove too faint or contaminated to justify observation. Hence we consider that the current census is approaching the limits of what is achievable with 8m-class telescopes and modern instrumentation under reasonable time constraints and, for the most evolved stars, is indeed essentially complete.

Finally, for completeness we note that the census does not sample the  additional population of $L-$band excess stars
which, via the presence of CO-bandhead emission, appear to support cool, dusty circumstellar 
discs (Stolte et al. \cite{stolte10}, \cite{stolte15}).

\subsection{Properties of the Arches stellar population}
 
We may summarise the current stellar census of the Arches as follows (Paper I and Table 1):
\begin{itemize}
\item{13 Wolf Rayets, comprising two WN7-8ha and eleven WN8-9ha subtypes.}
\item{Eight O hypergiants with spectral types ranging from O4-5 to O7-8.} 
\item{30 O supergiants spanning a restricted range of spectral types (O4 to O6).}
\item{Five lower luminosity O4-O6 I-III (super-)giants.}
\item{Ten O5-6.5 III giants.}
\item{Seven stars with spectral types ranging from O5-6  to O7-8 and of uncertain (III-V) luminosity class.}
\item{32 main sequence stars with spectral types spanning O5-6 V to O9.5 V.}
\end{itemize}

Following the discussion of completeness in Sect. 4.1, for any reasonable (initial) mass function (cf. Sect. 1) we are likely to be somewhat incomplete for the main sequence component of the Arches. Conversely, given the magnitudes of the  currently identified Wolf-Rayet and O hypergiants, we are likely to be essentially complete for both cohorts, and largely complete for the O supergiants; only three stars that are brighter than the current faintest O supergiant (F65) remain to be observed. 

The expanded census again emphasises the remarkable homogeneity of the cluster membership (cf. Paper I). Unlike the older Quintuplet ($\sim3-3.6$Myr) and Westerlund 1 ($\sim5$Myr) we find no examples of H-free WC stars within the Arches ($\sim2-3$Myr; Clark et al. \cite{clark05}, \cite{clark18b}, Crowther et al. \cite{crowther06}). We emphasise that the faintest WC stars within the Quintuplet - which derive from stars with initial masses comparable to those observed here - have $m_{\rm F205W}\sim11.7$ (Clark et al. \cite{clark18b}); a magnitude at which we are observationally  complete (Sect. 4.1). Likewise
the distribution of WN  spectral sub-types within the Arches is much narrower than seen in either of these clusters (Clark et al. \cite{clark18b}, Crowther et al. \cite{crowther06}). Similarly, we see no evidence for the earlier -  and highly luminous -  WN5-7ha and O2-3 Ia stars found within  younger clusters  such as NGC3603 ($\sim1-2$Myr; Melena et al. \cite{melena}, Roman-Lopes et al. \cite{roman}) and  R136 ($\sim1.5^{+0.3}_{-0.7}$Myr; Crowther et al. \cite{crowther16}). Finally, while we cannot exclude the possibility of massive blue stragglers within the cluster, we find no evidence to mandate their presence; this contrasts with the presence of demonstrably younger stellar members in both the Quintuplet and Wd1 (Clark et al. \cite{clark18b}, \cite{clark19}). 

The absence of the very earliest O spectral sub-types extends to the less-evolved stellar cohorts of the Arches. Our much expanded MS census suggests an absence of O2-4 stars, with the earliest stars present classified as O5-6 III-V or O5-6 V\footnote{F82 and F87, F90, F92, F102 and F115, respectively.}. Thus a MS turn-off around O4 V is still favoured; implying a surprisingly conservative turn-off mass of $\sim30-40M_{\odot}$ (following the {\em dynamically determined} spectral type/mass relation presented in Paper I). At the fainter end we are able to (re-)classify a large number of stars with previously low S/N spectra from $\geq$O8 V to $\sim$O9.5 V (noting that, in the absence of observational templates, our synthetic spectra 
suggest O8.5 V stars may be similar in appearance; Fig. 4). Therefore it would seem likely that the remaining stars with this preliminary classification (Paper I) are similarly only slightly later in spectral type. Again following Paper I, we estimate a mass of $\sim16M_{\odot}$ ($\sim22M_{\odot}$) for the O9.5 V (O8.5 V) stars. 

\subsection{Synopsis and future directions}

 The inclusion of data from the 2018 observing season has significantly expanded upon the stellar census  presented in Paper I,  
 allowing the first classifications of 18 stars and re-classification of a further 14; $\sim30$\% of a total of 105 classified 
 cluster  members (which does not include a further four deeply blended objects nor two stars with anomalous, unclassifiable 
 spectra: Sect. 3.1). Our results  buttress conclusions from Paper I; the Arches appears highly homogeneous and hence
 likely co-eval, with no evidence for H-free Wolf-Rayets nor the products of binary interaction, although we may not exclude the 
 presence of the latter at this time. Comparison to the cluster population suggests that three isolated, very massive 
 stars $\leq$5pc (projected)  from the Arches are potential runaways either stripped from the cluster by tidal forces or ejected 
 via dynamical interactions, with the nature of a fourth currently uncertain. 

 The enlarged and better-constrained MS population is arguably the most valuable product of this study. We find  an apparent  MS turn-off 
around O4-5 V, and classify  a large number of O9.5 V stars for the first time.   The initial masses of spectroscopically 
classified stars within the Arches therefore ranges from $\sim16M_{\odot}$ for the  O9.5 V stars through to $\geq120M_{\odot}$ 
 for the WN8-9ha primary of F2 (Lohr et al. \cite{lohr18a}). Filling in the  gaps and masses in the $\sim30-40M_{\odot}$ 
range are suggested for the O5-6 V cohort that delineates the MS turn-off (Sect. 4.2).  A {\em current} mass of $60\pm8M_{\odot}$ 
was determined for the O5-6 Ia$^{+}$ secondary of F2 (Lohr et al. \cite{lohr18a}); however  simulations by Groh et al. 
(\cite{groh14}) suggest that non-rotating stars of $60M_{\odot}$ do not pass through such  a phase, implying that mid-O hypergiants must 
instead evolve from stars of higher {\em initial} mass; a result supported by the independent  calculations  of Martins \& 
Palacios (\cite{mp}). Groh et al. (\cite{groh14}) further  demonstrate that in the 1.7-3Myr window   $60M_{\odot}$ stars will 
appear as, progressively, O4 Ia to O7 Ia supergiants; consistent  with both  the  estimated age of the  Arches and the 
distribution of supergiant spectral types (Paper I and Table 1).
 
  If correct, following from the population breakdown in Sect. 4.2 the Arches would host at least 51 stars with initial masses
 $\gtrsim60M_{\odot}$ (WRs and  O super-/hypergiants, but excluding the putative runaways from this count) with a subset likely 
 over twice this value. In contrast the Quintuplet hosts a  minimum of 38 stars with initial masses $\gtrsim60M_{\odot}$  (Clark 
 et al. \cite{clark18b}); however this is likely to be an underestimate since the current survey  is significantly incomplete and the 
 count does not include potential binary-interaction products nor the large population of H-free WC stars which do not conform to 
 the predictions of Groh et al. (\cite{groh14}).

The consequences of such a stellar cohort are striking. Within the next $\sim10$Myr, we might expect all 105 spectroscopically 
classified  members of the Arches to undergo core-collapse, leaving relativistic remnants behind (Groh et al. \cite{groh13}). 
The  incomplete nature of our current MS census - and indeed our insensitivity to stars $<15M_{\odot}$ - coupled with the 
possibility of a substantial population of massive binaries (Lohr et al. 
\cite{lohr19}) suggests that  this may be a significant underestimate of the true number. Evidently clusters such as the Arches and Quintuplet are 
important engines for the injection of neutron stars and black holes into 
the Galactic CMZ.

 Moving forward, the increased S/N of many of the Arches spectra will greatly aid in quantitative model-atmosphere analysis  even 
 where no spectral re-classification was warranted. Critically, the well defined observational properties of our expanded MS 
cohort will permit the simultaneous determination of an extinction law appropriate for the galactic centre; these efforts will be 
published in a future work. The numerous  dynamical  mass determinations available for O5-O9.5 V stars (Paper I) will inform 
calibration of the mass/luminosity function  for the Arches and hence the formulation of the cluster initial mass function. 
Likewise the physical parameters of the cluster  members determined via modelling will allow the construction of an HR diagram; 
allowing  current evolutionary predictions for  very massive stars to be tested and bulk cluster properties such as age and radiative and 
mechanical feedback to be constrained.

\begin{acknowledgements}
Based on observations collected at the European Organisation for Astronomical Research in the 
Southern Hemisphere under ESO programmes  087.D-0317, 091.D-0187, 093.D-0306, 099.D-0345 and 0101.D-0141. 
This research was supported by the Science and
Technology Facilities Council. FN acknowledges financial support through Spanish grants 
ESP2015-65597-C4-1-R and ESP2017-86582-C4-1-R (MINECO/FEDER).

\end{acknowledgements}

{}

\begin{thebibliography}{}

\bibitem[2000]{clark00}
Clark, J. S. \& Steele, I. A. 2000, A\&AS, 141, 65


\bibitem[2005]{clark05}
Clark, J. S., Negueruela, I., Crowther, P. A. \& Goodwin, S. P. 2005, A\&A, 434, 949

\bibitem[2018a]{clark18a}
Clark, J. S., Lohr, M. E., Najarro, F., Dong, H. \& Martins, F.
2018a, A\&A, 617, A66

\bibitem[2018b]{clark18b}
Clark, J. S., Lohr, M. E., Patrick, L. R., et al. 2018b, A\&A, 618, A2

\bibitem[2018c]{clark18c}
Clark, J. S., Lohr, M. E., Najarro, F.  et al. 2018c, The Messenger, 173, 22

\bibitem[2019]{clark19}
Clark, J. S., Najarro, F., Negueruela, I. et al. 2019, A\&A, in press (arXiv:1811.09444)

\bibitem[2012]{clarkson}
Clarkson, W. I., Ghez, A. M., Morris, M. R. et al. 2012, ApJ, 751, 132


\bibitem[1996]{cotera96}
Cotera, A. S., Erickson, E. F., Colgan, S. W. J. et al.  1996, ApJ, 461, 750

\bibitem[1999]{cotera99}
Cotera, A. S., Simpson, J. P., Erickson, E. F. et al. 1999, ApJ, 510, 747

\bibitem[2006]{crowther06}
Crowther, P. A., Hadfield, L. J., Clark, J. S., Negueruela, I. \& Vacca, W.
D. 2006, MNRAS, 372, 1407




\bibitem[2016]{crowther16}
Crowther, P. A., Caballero-Nieves, S. M., Bostroem, K. A., et al. 
2016, MNRAS, 458, 624



\bibitem[2014]{demink}
de Mink, S. E., Sana, H., Langer, N., Izzard, R. G. \& Schneider, F. R N.
2014, ApJ, 782, 7

\bibitem[2011]{dong11}
Dong, H., Wang, Q. D.,  Cotera, A. et al.  2011, MNRAS, 417, 114

\bibitem[2015]{dong15}
Dong, H., Mauerhan, J., Morris, M., Wang, Q. D. \& Cotera, A. 2015, MNRAS, 446, 842

\bibitem[2009]{espinoza}
Espinoza, P., Selman, F. J. \& Melnick, J. 2009, A\&A, 501, 563


\bibitem[1999]{figer99}
Figer, D. F.,  Kim, S. S., Morris, M., et al. 1999, ApJ, 525, 750



\bibitem[2002]{figer02}
Figer, D. F., Najarro, F., Gilmore, D. et al. 2002, ApJ, 581, 258

\bibitem[2013]{groh13}
Groh, J. H., Meynet, G., Georgy, C. \& Ekstr\'{o}m, S.  2013, A\&A, 558,
A131


\bibitem[2014]{groh14}
Groh, J. H., Meynet, G., Ekstr\'{o}m, S. \&  Georgy, C. 2014, A\&A, 564, 
A30


\bibitem[2013]{habibi}
Habibi, M., Stolte, A., Brandner, W. Hu{\ss}man, B. \& Motohara, K. 
2013, A\&A, 556, A26

\bibitem[2014]{habibi14}
Habibi, M., Stolte, A. \& Harfst, S. 2014, A\&A, 566, A6

\bibitem[2005]{hanson05}
Hanson, M. M., Kudritzki, R.-P., Kenworthy, M. A., Puls, J. \&
Tokunaga, A. T. 2005, ApJS, 161, 154 

\bibitem[1998]{hm1}
Hillier, D. J. \& Miller, D. L. 1998, ApJ, 496, 407

\bibitem[1999]{hm2}
Hillier, D. J. \& Miller, D. L. 1999, ApJ, 519, 354

\bibitem[2018]{hosek}
Hosek, M. W., Lu, J. R., Anderson, J. et al. 2018, ApJ, in press (arXiv:1808.02577)


\bibitem[2006]{kim}
Kim, S. S., Figer, D. F., Kudritzki, R. P. \& Najarro, F.
2006, ApJ, 653, L113

\bibitem[2018]{lohr18a}
Lohr, M. E., Clark, J. S., Najarro, F. et al. 2018, A\&A, 617, A66

\bibitem[2019]{lohr19}
Lohr, M. E., Clark, J. S., Patrick, L. R. et al. 2019, in prep.


\bibitem[2005]{martins05}
Martins, F., Schaerer, D. \& Hillier, D. J. 2005, A\&A, 436, 1049

\bibitem[2008]{martins08}
Martins, F., Hillier, D. J., Paumard, T. et al. 2008, A\&A, 478, 219

\bibitem[2017]{mp}
Martins, F. \& Palacios, A. 2017, A\&A, 598, A56


\bibitem[2010]{mauerhan}
Mauerhan, J. C., Cotera, A., Dong, H. et al. 2010, ApJ, 725, 188 


\bibitem[2008]{melena}
Melena, N. W., Massey, P., Morrell, N. I. \& Zangari, A. M. 2008, AJ, 135, 878



\bibitem[1995]{nagata}
Nagata, T., Woodward, C. E., Shure, M. \& Kobayashi, N. 1995, AJ, 109, 1676


\bibitem[2004]{najarro}
Najarro, F., Figer, D. F., Hillier, D. J. \&  Kudritzki, R. P. 2004, ApJ, 611, L105

\bibitem[2009]{paco09}
Najarro, F., Figer, D. F., Hillier, D. J., Geballe, T. R. \& Kudritzki, R. P.
2009, ApJ, 691, 1816


\bibitem[2018]{park}
Park, S.-M., Goodwin , S. P. \& Kim, S. S. 2018, MNRAS, 478, 183

\bibitem[1967]{poveda}
Poveda, A., Ruis, J. \& Allen, C. 1967, BOTT, 4, 86

\bibitem[2016]{roman}
Roman-Lopes, A., Franco, G. A. P. \& Sanmartim, D.    2016, ApJ, 823, 96 

\bibitem[2014]{schneider}
Schneider, F. R. N., Izzard, R. G., de Mink, S. E., et al. 2014, ApJ, 780, 117

\bibitem[2015]{shin}
Shin, J. \& Kim, S. S. 2015, MNRAS, 477, 366

\bibitem[2002]{stolte02}
Stolte, A., Grebel, E. K., Brandner, W. \& Figer, D. F. 2002, A\&A, 394, 
459 

\bibitem[2010]{stolte10}
Stolte, A., Morris, M. R., Ghez, A. M., et al. 
2010, ApJ, 718, 810

\bibitem[2015]{stolte15}
Stolte, A., Hu{\ss}mann, B., Olczak, C., et al. 2015, A\&A, 578, A4

\end{thebibliography}
\end{document}